# A systematic study to investigate the effects of X-ray exposure on electrical properties of silicon dioxide thin films using X-ray photoelectron spectroscopy

Carlos Munoz, Thomas Iken, Nuri Oncel†

Department of Physics & Astrophysics, University of North Dakota, Grand Forks, North Dakota, 58202, USA

†Author to whom correspondence should be addressed: nuri.oncel@und.edu

**Abstract:**

X-ray Photoelectron Spectroscopy (XPS) is generally used for chemical analysis of surfaces and interfaces. This method involves the analysis of changes in binding energies and peak shapes of elements under consideration. It is also possible to use XPS to study the effect of X-ray radiation on the electrical properties of thin films. We measured the Si 2p peak using X-ray powers of 300 W and 150 W on approximately 135 nm silicon dioxide ($SiO_2$) thin films grown on both n- and p-type substrates while applying DC or AC external biases. Using the shifts in the binding energy of the Si 2p peak, we calculated the resistances and the capacitances of the $SiO_2$ thin film. The way that the binding energies of the Si 2p peak and the capacitance of the thin film change as a function of the type of Si substrate and the power of the X-ray are explained using band bending.

**Introduction:**

X-ray Photoelectron Spectroscopy (XPS) is an important tool in the analysis of the chemical and physical properties of materials and interfaces.[1,2] For insulating samples, the grounding of the sample may not be sufficient to prevent the accumulation of positive charges on the surface. This leads to differential charging, which for a long time, has been considered an issue. Electron flood guns have been used to control and reduce this effect, but eliminating differential charging is difficult, and if not done properly can lead to erroneous results. However, a useful aspect of differential charging is it can be employed to study the structural and electronic properties of dielectric films, where equivalent circuit models have been proposed and employed to extract the capacitance and resistance from the observed shifts in binding energy with applied external AC or DC voltage[3–11]. Tasci et al. developed computer-controlled electronics for capacitance measurements [5]. However, this method is limited by dwell time, the amount of time needed to collect photoelectrons at a given energy. The dwell time cannot be randomly small and needs to be long enough to have a meaningful signal-to-noise ratio. To avoid limitations due to dwell time, we derived closed-form mathematical equations to describe binding energy shifts as a function of applied bias. This relatively simpler approach allowed us to use a curve-fitting algorithm to extract the resistance and capacitance of a thin film of Si nanocrystals grown on a polycrystalline Ag(111) surface [3]. The method we propose has virtually no limitation on the time constant of the RC circuit.

It is important to remember that XPS inherently utilizes an X-ray source, and the effect of X-ray radiation can be significant on the electrical properties of the thin films [12–17]. In a typical XPS paper, the focus is on the changes in the binding energy and associated surface chemistry, neglecting the fact that X-rays easily penetrate a few microns into the sample and damage the sample. A typical XPS analysis would not be sufficient to study these effects. However, by applying AC/DC biases, it is possible to investigate the effect of X-rays on the substrate and interfaces buried deep under the surface. The $SiO_2$/Si(100) system has been studied countless times and it is especially important for semiconductor technology. However, to the best of our knowledge, the effect of X-ray power on the electronic properties of the system has not been studied systematically by using similarly doped n-type and p-type samples. Therefore, in this paper, we performed a systematic study of $SiO_2$ thin films grown on n-type and p-type Si(100) samples using 150 W and 300 W X-ray power to study how the variation of X-ray power affects the electrical properties of $SiO_2$ thin films grown on Si samples that are doped differently.

**Experimental Methods**

P-type and n-type Si(100) wafers with a resistivity of $0.1 - 1$ Ω cm were chosen for the preparation of the experimental circuit seen in Figure 1a. Wafers were cut to a size of 15 mm × 8mm × 0.5 mm. A Thermolyne 46100 high-temperature furnace was used to grow $133 \pm 2$ nm thick oxide layers using a dry oxygen method at 1100 °C. A Lambda Scientific LEOI-44 ellipsometer was employed to verify the thickness of the oxide layers. The XPS utilized for this study has a PHI Model 5400 electron spectrometer with a non-monochromatized Mg Kα (1253.6 eV) as an X-ray source. During XPS measurements, the chamber pressure was kept at or below $1 \times 10^{-9}$ mbar. All XPS core-level spectra were analyzed using Auger scan software equipped with its curve-fitting program. The core-level peaks were fitted using a Gaussian–Lorentzian (GL) function to include the instrumental response function along with the core-level line shape. The secondary-electron background was subtracted using a Shirley function. A thin sheet of Au foil was attached to the edge of each sample and grounded directly via the sample holder. In our study, we used two sets of samples. For the first set of samples, the X-ray power was set at 300 W. When the samples were grounded, the binding energy Si 2p peaks were measured at 104.79 eV (n-type) and 104.61 eV (p-type). For the second set of samples, the X-ray power was set at 150 W. The binding energy of the Si 2p peak for the n-type sample remained about the same at 104.74 eV at this power. On the other hand, for the p-type sample, the binding energy of the Si 2p peak shifted to a lower value of 104.29 eV. An Agilent 33500B waveform generator was used as a function generator. We applied a 10 V DC bias and variable frequency 0–10 V square wave pulses to determine the electrical characteristics of the $SiO_2$ thin films. The duty cycle of the square waves was 50% with frequencies varying from 400 µHz to 1 KHz. Raw data from scans was smoothed in the ORIGIN suite using a 10pt LOWESS (locally weighted scatter plot smoothing) algorithm. To calculate the fitting of the experimental data, we used the NonlinearModelFit function in Wolfram Mathematica software.

**Results and Discussion:**

The positions of XPS peaks are measured relative to the Fermi level of the substrate. For doped semiconductors, dopant-induced states determine the position of the bulk Fermi level that leads to measurable binding energy shifts with respect to the intrinsic Fermi level of the pristine substrate. A calculation using the conductivities provided by the manufacturer of the Si samples showed that a 0.68 eV binding energy difference should exist between the Si 2p of n-type and p-type samples. (See Supplementary Material)

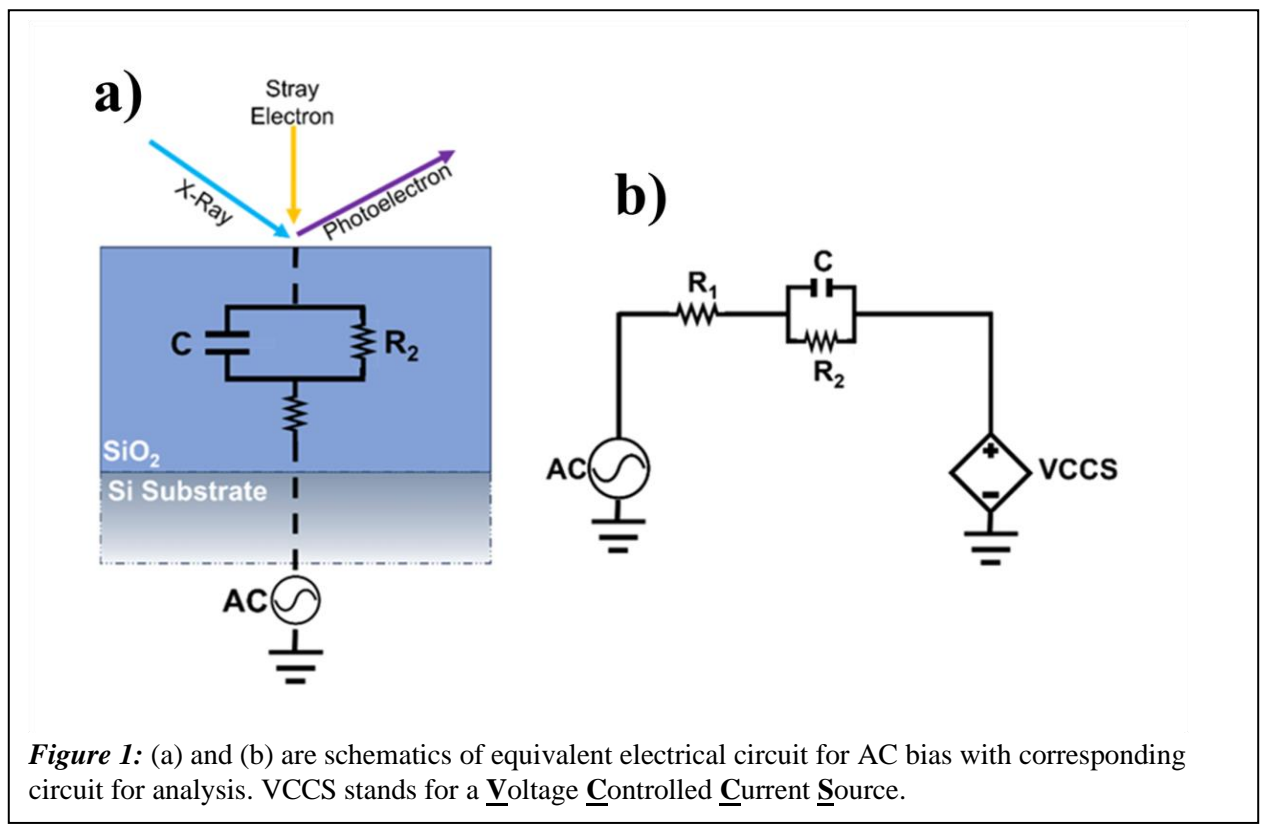

*Figure 1:* (a) and (b) are schematics of equivalent electrical circuit for AC bias with corresponding circuit for analysis. VCCS stands for a **V**oltage **C**ontrolled **C**urrent **S**ource.

However, only 0.18 eV binding energy differences were measured when 300 W X-ray power was used. This difference is attributed to the band bending caused by the difference between the Fermi levels of the analyzer and the samples[18]. This means that for both n-type and p-type samples a depletion layer has formed (See Supplementary Material). In the $SiO_2$ layer, X-ray radiation creates electron-hole pairs. Compared to holes, electrons escape more easily from the recombination process due to their higher mobility. The holes can be trapped in the oxide layer or at the interface between oxide and Si. Therefore, the net oxide trapped charges are expected to be positive [19]. The effect of the (+) trapped charges on the depletion layer of n-type and p-type are opposite. On a p-type sample, the band bending due to (+) trapped charges and the Fermi level difference are additive whereas on an n-type sample, they are not [20]. To investigate further the effect of the X-ray intensity on band bending, we repeated the measurements with 150 W X-ray power. Using 150 W X-ray power, the binding energy difference of Si 2p increased to 0.45 eV getting closer to the theoretically predicted 0.68 eV confirming that X-ray alters the band bending (Table 1). The binding energy change is more prominent on the p-type of the sample. This can be attributed to the (+) trapped charges in the oxide layer. When the power of the X-rays is reduced, so are the number of trapped (+) charges in the oxide layers, effectively reducing the band bending at the interface and leading to an increased binding energy difference.

| Si 2p Binding Energy | n-type (150 W) | n-type (300 W) | p-type (150 W) | p-type (300 W) |
|---|---|---|---|---|
| Grounded | 104.74 eV | 104.79 eV | 104.29 eV | 104.61 eV |

**Table 1:** Grounded Si 2p binding energy

To simulate the electrical response of our system, we used an equivalent circuit model for a real capacitor, using two resistances – one in series and one in parallel with our capacitor [3] (See Figure 1b). In an ideal capacitor, electrical energy is stored and released without loss. However, real capacitors have defects that can create internal resistances and leakage currents. As seen in Figure 1b, $R_1$ represents the resistor in series and $R_2$ represents the resistor in parallel which also carries a leakage current. Electrons flow from the sample surface and free electrons in the vacuum are represented by the voltage-controlled current source (VCCS) [21]. To extract the capacitance of our thin films, we performed a sequence of XPS measurements with 10 V square waves at frequencies ranging from 400 µHz to 1KHz. In all cases, the shift in binding energy exhibits a sigmoidal frequency dependence, characteristic of RC circuits under AC power (Figure 2). We assumed that the currents flowing through the samples remain constant in the dynamic range. In our previous work, we used three different models to extract the resistance and capacitance of a thin film of Si nanocrystals. Results obtained using all three models agreed well with each other within the standard error of each model. Therefore, for the current system, we decided to use Eq.1 to fit the experimental data and determine $R_1$, $R_2$, and $C$ (Table-2, 1st row). [Eq.1]. Eq.1 represents a full solution of the proposed circuit for an AC source at a fixed frequency but avoids the unnecessary complications of the Fourier series used to construct square wave pulses as harmonically related sinusoids. The results of the fitting and raw XPS data can be seen in Figure 2.

$$\Delta V_\omega = 10\ \text{V} - I\left(R_1 + \frac{1}{\sqrt{\frac{1}{R_2^2}+(\omega C)^2}}\right) \tag{1}$$

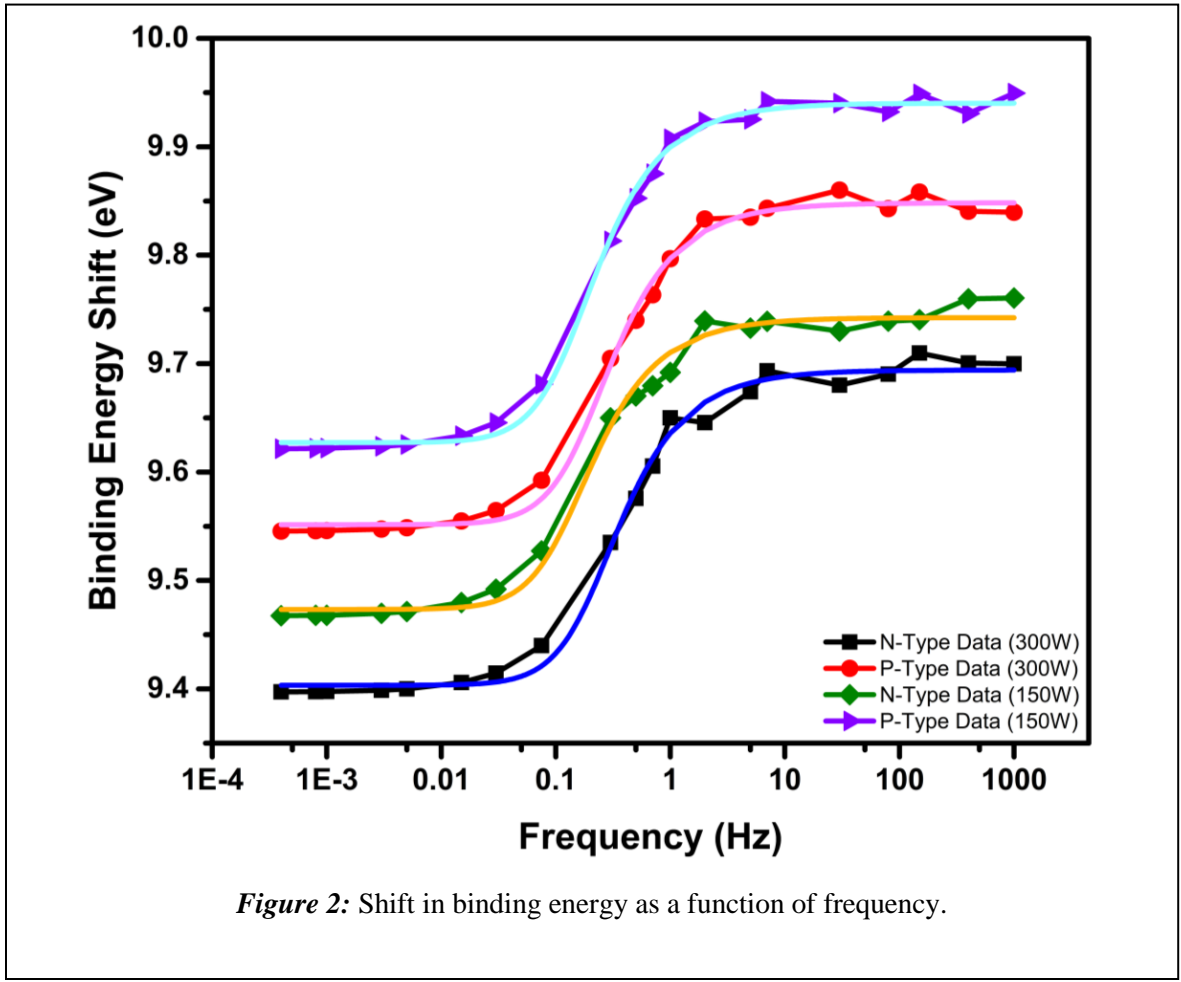

*Figure 2:* Shift in binding energy as a function of frequency.

A thermally grown silicon oxide thin film has a relative dielectric constant of 3.9 and resistivity of ~ $10^{12}$ $\Omega \cdot$ cm. [22] A back-of-an-envelope calculation would show that the capacitance and resistance of the 133 nm film should be approximately 31 nF and 11 M$\Omega$, respectively. A comparison of these numbers with the values in Table 2 for capacitance and equivalent resistance ($R_{eq}$) shows that XPS overshoots them both. In addition, when the power of the X-ray was reduced from 300 W to 150 W, the capacitance of $SiO_2$ grown on p-type samples decreased significantly, whereas the capacitance of the $SiO_2$ on n-type samples remained about the same. A sample under X-ray exposure can be considered a typical MOS device with an effective (+) gate voltage. In our study we found decreasing the intensity of incident X-rays has two effects. First, it lowers the applied effective (+) gate bias and second, it decreases the number of trapped (+) charges in the oxide layer. Both effects work together to lower the (-) charge accumulated at the $SiO_2$/Si interface causing a resultant decrease in the capacitance ($C = \frac{Q}{V}$). These effects are overall more pronounced in the p-type samples because of the additive nature of band bending inherit to p-type semiconductors and the relatively higher number of ionized dopants at the $SiO_2$/Si interface which can more effectively respond to changes. Additionally, the resistances of $SiO_2$ grown on both n-type and p-type samples exhibited a substantial increase when the X-ray power was reduced by 50%. This observation is in line with MOS devices exposed to radiation. Yilmaz et al. observed that the current transport through the $SiO_2$ film increases due to the easier conduction path formed through radiation-induced damage in the oxide layer [19].

|  | expected | p-type (300 W) | p-type (150 W) | n-type (300 W) | n-type (150 W) |
|---|---|---|---|---|---|
| C (nF) | 31 | 76.47 ± 5.01 | 48.85 ± 2.88 | 67.01 ± 4.96 | 61.43 ± 6.30 |
| $R_1$ |  | 6.06 ± 0.13 | 4.78 ± 0.22 | 12.24 ± 0.15 | 20.61 ± 0.32 |
| $R_2$ |  | 11.87 ± 0.18 | 25.04 ± 0.31 | 11.63 ± 0.21 | 21.53 ± 0.45 |
| $R_{eq}$ (M$\Omega$) | 11 | 17.93 ± .31 | 29.82 ± .53 | 23.86 ± .36 | 42.13 ± .77 |

**Table 2:** Resistance and capacitance values extracted by fitting the experimental data with Eq. 1.

**Conclusion:**

Most of the time, when measuring XPS the effect of X-ray on physical and chemical properties of samples is ignored. For a sample as robust as $SiO_2$, we showed that X-ray radiation changes the electrical properties of the thin film by applying AC/DC biases while performing standard XPS analysis. The XPS+$SiO_2$/Si system can be considered a typical MOS capacitor exposed to radiation. Due to the attachment of the sample to the analyzer and the work function difference between them, significant band bending happens at the interface. This is evident in how resistance and capacitance values of the $SiO_2$ thin films show a strong dependency on X-ray power. Although the X-ray radiation strengthens the band bending on p-type samples, it has the opposite effect on n-type samples.

**Supplementary Material**

The supplementary material provides a calculation of the positions of the Fermi levels for n-type and p-type samples. Figure 1 presents XPS data showing how the binding energy difference increased as the intensity of the X-ray beam was reduced. Figure 2 is the energy band diagram of the samples and analyzer.

**Conflict of Interest:**

The authors have no conflicts to disclose.